\documentclass[aps,twocolumn,prl,showpacs,floatfix,altaffilletter]{revtex4-1}
\usepackage{graphicx}
\usepackage{bm}
\usepackage{amsmath}
\usepackage{sansmath}
\usepackage{amsfonts}
\usepackage{amssymb}
\usepackage{xspace}
\usepackage{xr}
\usepackage{natbib} %\bibliographystyle{prsty}
\newcommand{\beq}{\begin{equation}}
\newcommand{\eeq}{\end{equation}}

\begin{document}

\title{Neutron Scattering Signatures of Magnon Weyl Points}

\author{S.~Shivam$^{1,2}$, R.~Coldea$^3$, R. Moessner$^{1}$, P.~A.\ McClarty$^{1}$ }

\affiliation{$^{1}$ Max Planck Institute for the Physics of Complex Systems, N\"{o}thnitzer Str. 38, 01187 Dresden, Germany}
\affiliation{$^{2}$ Department of Physics, Indian Institute of Technology Bombay, Mumbai 400076, India}
\affiliation{$^{3}$ Clarendon Laboratory, University of Oxford, Parks Road, Oxford, OX1 3PU, UK}

%\pacs{75.10.Jm, 75.30.Ds,75.25.-j}

% 75.10.Jm Quantized spin models, including quantum spin frustration
% 75.30.Ds Spin waves (for spin-wave resonance, see 76.50.+g)

\begin{abstract}
We study the inelastic neutron scattering cross section in the vicinity of touching points in magnon bands. Among the possible touching points are magnon Weyl points in three dimensional ordered magnets with significant spin-orbit coupling that are characterized by a linear dispersion in their vicinity. A Weyl point is topologically protected by its net chirality and here we show that this leads to a characteristic form for the dynamical structure factor. To address this question, we show that scattering intensities in the vicinity of arbitrary magnon two-band touching points are identical to expectation values of the pseudospin polarization along some direction $\hat{\boldsymbol{n}}$ in momentum space fixed by the magnetic Hamiltonian. This approach applied to Weyl points shows that they are singular points in the intensity of the form $\hat{\boldsymbol{n}}\cdot \delta\hat{\boldsymbol{k}}$ regardless of the magnetic ground state. We make specific predictions for the experimental signatures of such intensity singularities in several spin models hosting Weyl magnons applicable to candidate materials.
\end{abstract}

\maketitle

%%%%%%%%%%%%%%%%%%%%%%%%%%%%%%%%%%%%%%%%%%%%%%%%%%%%%%%%%%%%%
%%%%%%%%%%%%%%%%%%%%%%%%%%%%%%%%%%%%%%%%%%%%%%%%%%%%%%%%%%%%%
%% INTRODUCTION
%%%%%%%%%%%%%%%%%%%%%%%%%%%%%%%%%%%%%%%%%%%%%%%%%%%%%%%%%%%%%
%%%%%%%%%%%%%%%%%%%%%%%%%%%%%%%%%%%%%%%%%%%%%%%%%%%%%%%%%%%%%

Weyl semimetals are solids in which the valence and conduction bands meet at points with linear dispersion in the three dimensions of momentum space, the chemical potential being located at the touching point \cite{wan2011topological}. Weyl points are monopoles in the Berry curvature associated with a nonzero first Chern number on closed surfaces enclosing them. Such band structures have been found in a number of materials \cite{xu2015discovery,xu2015experimental,arnold2016negative,lv2015experimental,lv2015observation,weng2015weyl,huang2015weyl,shekhar2015extremely,yang2015weyl}.  The presence of Weyl points in the band structure has various observable consequences.  For example, the presence of a boundary enforces surface states that are gapless on open curves on the surface Brillouin zone - so-called Fermi arcs \cite{wan2011topological}. The bulk magnetotransport properties are also affected by the presence of Weyl points through the chiral anomaly \cite{zyuzin2012topological,son2013chiral}. 

Weyl points may also appear in a very different context: in the magnon bands of materials with long-range magnetic order. In cases where the single particle description of the magnon bands is a good approximation, Weyl points are topologically protected and there are well-defined gapped surface states - the magnon analogue of Fermi arcs. To date, a number of physically motivated spin models  exhibiting Weyl points in their magnon spectra have been theoretically proposed \cite{li2016weyl,mook2016tunable,su2017magnonic,su2017chiral,owerre2017weyl} providing compelling grounds that they will be observed in real magnetic materials. The most suitable tool to explore the bulk magnon band structure is inelastic neutron scattering (INS) that provides an energy-momentum resolved map of the magnon dispersion relations.

Just as electronic Weyl points have experimental signatures arising from the electronic wavefunction and the band topology, it is natural to ask whether magnon Weyl point wavefunctions produce a characteristic pattern in the INS intensity modulations.
In this paper, we present a general framework for thinking about these intensity patterns that is applicable to two-band touching points. We find that the neutron cross section about a touching point in momentum space can be recast as the problem of measuring the polarization of momentum dependent spinors along some non-universal direction characteristic of the interaction Hamiltonian near the touching point.  We apply the method to various examples including Dirac points in 2D magnets, Weyl points in 3D magnets and linear-quadratic touching points. We propose specific experimental signatures for Weyl points - for which the spinors form a hedgehog texture around the point - so the intensity has an anisotropy axis along which the intensity flips from maximum to minimum passing though the Weyl node.

%%%%%%%%%%%%%%%%%%%%%%%%%%%%%%%%%%%%%%%%%%%%%%%%%%%%%%%%%%%%%
%%%%%%%%%%%%%%%%%%%%%%%%%%%%%%%%%%%%%%%%%%%%%%%%%%%%%%%%%%%%%
%% SPIN WAVES AND INELASTIC NEUTRON CROSS SECTION
%%%%%%%%%%%%%%%%%%%%%%%%%%%%%%%%%%%%%%%%%%%%%%%%%%%%%%%%%%%%%
%%%%%%%%%%%%%%%%%%%%%%%%%%%%%%%%%%%%%%%%%%%%%%%%%%%%%%%%%%%%%

To be concrete, we concentrate on the case of scattering from insulating magnets. The differential neutron scattering cross section measured by INS is proportional to the correlation function $\frac{d^2 \sigma}{dE d\Omega} \propto \sum_{\alpha\beta} \left( \delta_{\alpha\beta} - \hat{k}^{\alpha}\hat{k}^{\beta} \right) S^{\alpha\beta}(\boldsymbol{k},\omega)$ where \cite{squires_2012} 
\begin{equation}
S^{\alpha\beta}(\boldsymbol{k},\omega) = -\frac{1}{\pi} {\rm Im} \int_{-\infty}^{\infty} dt e^{-i\omega t} \mathcal{G}^{\alpha\beta}(\boldsymbol{k},\omega)
\end{equation}
and $\boldsymbol{k}$ and $\omega$ are, respectively, the neutron momentum transfer and energy transfer and the Green's function $\mathcal{G}^{\alpha\beta}(\boldsymbol{k},\omega)= -i \left\langle {\cal T} M_{\boldsymbol{k}}^{\alpha}(t) M_{-\boldsymbol{k}}^{\beta}(0)  \right\rangle$. Here $M_{\boldsymbol{k}}^{\alpha}= \sum_{i,a} \mathsf{J}^{\alpha}_{ia} \exp\left( -i\boldsymbol{k}\cdot (\boldsymbol{R}_i + \boldsymbol{r}_a) \right)$ is the Fourier transformed moment. We suppose that magnetic unit cell has $N$ sites. The physical moments $\mathsf{J}^{\alpha}_{ia}$ live on lattice sites indexed by magnetic primitive lattice vector $\boldsymbol{R}_i$ and magnetic basis labelled by $a=1,\ldots,N$. We allow for the possibility that these moments have a local anisotropy parametrized by a g-tensor that is diagonal in the local frame obtained by rotation $R_a^{\alpha\beta}$ with moments labelled a tilde: $\mathsf{J}^{\alpha}_{ia}=R_a^{\alpha\beta} g^{\beta}\tilde{\mathsf{S}}_{ia}^{\beta}$. We compute the cross section within linear $1/S$ spin wave theory \cite{holstein1940field,SM} thereby neglecting interactions between the magnons.

One can show that
\begin{equation}
S^{\alpha\beta}(\boldsymbol{k},\omega) = \frac{S}{2} \sum_{m}\sum_{a,b}\sum_{\sigma,\tau =x,y} \Gamma_a^{\alpha \sigma} \Gamma_b^{\beta \tau} Q^{\sigma\star}_{\boldsymbol{k}am} Q^{\tau}_{\boldsymbol{k}bm}\delta(\omega - \epsilon_{\boldsymbol{k},m})
\end{equation}
where the spin wave energies are given by $\epsilon_{\boldsymbol{k},m}$ for mode $m$. The symbols entering into the structure factor are defined in the Supplementary Material \cite{SM}.

%%%%%%%%%%%%%%%%%%%%%%%%%%%%%%%%%%%%%%%%%%%%%%%%%%%%%%%%%%%%%
%%%%%%%%%%%%%%%%%%%%%%%%%%%%%%%%%%%%%%%%%%%%%%%%%%%%%%%%%%%%%
%% DIRAC POINT BY WAY OF ILLUSTRATION
%%%%%%%%%%%%%%%%%%%%%%%%%%%%%%%%%%%%%%%%%%%%%%%%%%%%%%%%%%%%%
%%%%%%%%%%%%%%%%%%%%%%%%%%%%%%%%%%%%%%%%%%%%%%%%%%%%%%%%%%%%%

\begin{figure}[ht!]
\begin{center}
\includegraphics[width=0.5\textwidth]{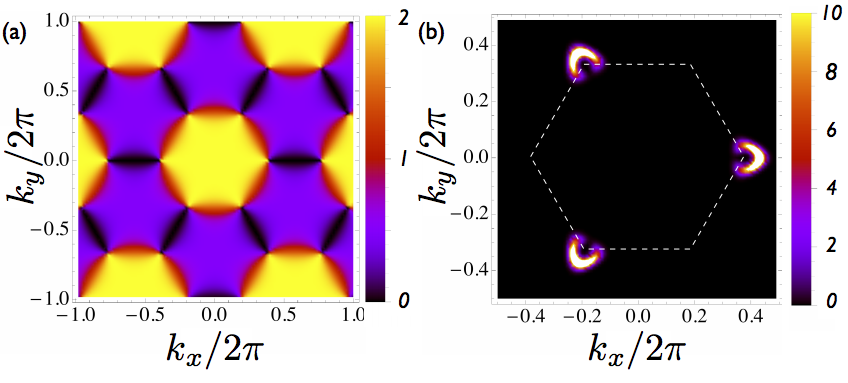}
\caption{(a) Dynamical structure factor in the lower magnon band of the isotropic honeycomb ferromagnet (Eq.~\ref{eq:Dirac}) showing the characteristic anisotropic intensity pattern around the Dirac points in the magnon spectrum at the zone corner K points. For the upper band the color scale is reversed. (b) Constant energy intensity map for the stacked honeycomb ferromagnet (see main text) showing ``intensity arcs" around the Weyl points which are located at the centres of the arcs. The calculation is performed for $L=1/4$ and energy $\omega/2JS=3.1$ compared to Weyl point energy of $\omega_0 /2JS=2.9$. The intensity arcs appear inverted below the Weyl node energy. The dashed hexagon indicates the Brillouin zone.
 }\label{fig:dirac_weyl}
\end{center}
%\vspace{-25pt}
\end{figure}

{\it Dirac Points in 2D Magnets} $-$ The main findings of this paper are perhaps best exemplified by the case of Dirac magnon touching points in a two band model. A collinear ferromagnet with nearest neighbor exchange and Hamiltonian $H=-\vert J \vert\sum_{\langle i,j\rangle} \boldsymbol{\mathsf{S}}_i\cdot\boldsymbol{\mathsf{S}}_j$ has a magnon band structure identical to the single orbital tight binding model on the same lattice. It follows that the honeycomb ferromagnet has the same magnon band structure as graphene with Dirac touching points at $K$ and $K'$. The spin wave Hamiltonian takes the form $\boldsymbol{\mathcal{H}} (\boldsymbol{k})=3\vert J\vert S \boldsymbol{\sigma}^0 - \vert J\vert S \boldsymbol{\sigma}^1 {\rm Re}[\gamma_{\boldsymbol{k}}]+\vert J\vert S \boldsymbol{\sigma}^2 {\rm Im}[\gamma_{\boldsymbol{k}}]$ where $\gamma_{\boldsymbol{k}}=\sum_{\mu=1,2,3} \exp(i\boldsymbol{k}\cdot\boldsymbol{\delta}_{\mu})$, the $\boldsymbol{\delta}_{\mu}$ are the three nearest neighbor bond direction vectors and the $\boldsymbol{\sigma}^\mu$ are the Pauli matrices. From the eigenvectors of the Hamiltonian, which may be diagonalized unitarily because there are no anomalous magnon pair creation terms, one may show that the dynamical structure factor of the magnons at wavevector $\boldsymbol{k}$ is
\begin{align}
S^{xx}(\boldsymbol{k}) = S^{yy}(\boldsymbol{k})  = \frac{S}{2}\left( 1\pm \frac{{\rm Re} [ \gamma_{\boldsymbol{k}} ]}{\vert \gamma_{\boldsymbol{k}} \vert} \right) 
\label{eq:Dirac}
\end{align}
with $+$ for the lower band and $-$ for the upper band. Fig.~\ref{fig:dirac_weyl}(a) shows the intensity map of the lower magnon band in several Brillouin zones assuming the spin ordering direction ($z$) is normal to the honeycomb plane. At the zone corners where the Dirac points are located the intensity varies smoothly on a contour surrounding each point while the point itself is singular. By expanding $\gamma_{\boldsymbol{k}}$ about each point it is straightforward to see that the intensity varies as $1\pm \cos\theta$ where the angle $\theta$ winds around the Dirac point with origin along the line joining $\Gamma$ to the chosen $K$. The sum over both bands has no singularities. 

The calculation presented above shows that the pseudospin winding around a magnon Dirac point has observable consequences in the neutron scattering cross section reminscent of ARPES scans of electronic Dirac materials \cite{mucha2008characterization}. In the following we uncover a generalization of this result to arbitrary two magnon band touchings.

\begin{figure*}[ht!]
\begin{center}
\includegraphics[width=2.1\columnwidth]{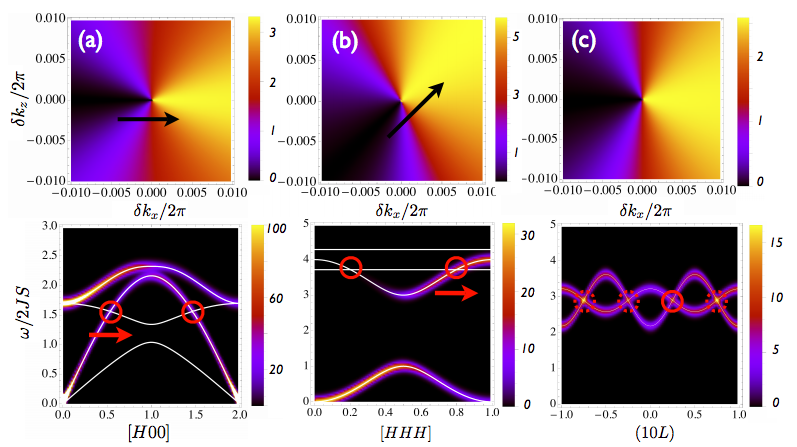}
\caption{Figures show the dynamical structure factor for three different models. (a) Breathing pyrochlore with Weyl point at $(0.535,0,0)$.  (b) Pyrochlore ferromagnet with Dzyaloshinskii-Moriya exchange with polarization along $\langle 111\rangle$ with Weyl point at $(x,x,x)$ with $x=0.798$. (c) Stacked honeycomb ferromagnet with DM exchange with Weyl point at $(1,0,1/4)$. The upper panels show the intensity winding around the Weyl point in the higher energy mode (a) in the $[HLL]$ plane, (b) in the $[HHL]$ plane and (c) the $(H0L)$ plane where $\delta\boldsymbol{k}$ components are measured with respect to the respective Weyl points. The lower panels show the dynamical structure factor along lines in momentum space with the dispersions overlaid. Circles mark the locations of the Weyl points and the arrows (top and bottom panels) show the anisotropy axes in the intensity. For the lower panel of (c), the anisotropy axis is perpendicular to the cut shown. 
 }\label{fig:weyl}
\end{center}
%\vspace{-25pt}
\end{figure*}

%%%%%%%%%%%%%%%%%%%%%%%%%%%%%%%%%%%%%%%%%%%%%%%%%%%%%%%%%%%%%
%%%%%%%%%%%%%%%%%%%%%%%%%%%%%%%%%%%%%%%%%%%%%%%%%%%%%%%%%%%%%
%% GENERAL TWO BAND TOUCHING PROBLEM AND PROJECTION OPERATOR
%%%%%%%%%%%%%%%%%%%%%%%%%%%%%%%%%%%%%%%%%%%%%%%%%%%%%%%%%%%%%
%%%%%%%%%%%%%%%%%%%%%%%%%%%%%%%%%%%%%%%%%%%%%%%%%%%%%%%%%%%%%

{\it General Two Band Touching} $-$ Consider the vicinity of a two band touching point in an $N$ magnon band model $\boldsymbol{\mathcal{H}} = \boldsymbol{\mathcal{H}}^{(0)} + \boldsymbol{\mathcal{H}}^{(1)}$, where $\boldsymbol{\mathcal{H}}^{(0)}$ is the Hamiltonian at the degenerate point with energy $\omega_0$ at wavevector $\boldsymbol{k}_{\rm W}$ and $\boldsymbol{\mathcal{H}}^{(1)}$ is the perturbation at wavevector $\boldsymbol{k}_{\rm W}+\delta\boldsymbol{k}$. The effective Hamiltonian operating within the subspace of the touching bands, denoted $n$ and $n+1$ may be obtained by projecting onto these bands and carrying out degenerate perturbation theory. If $\boldsymbol{v}^{(m)}_{\boldsymbol{k}_{\rm W}}$ are the appropriately orthonormalized magnon wavefunctions at the degenerate point for $m=n,n+1$ the effective Hamiltonian $H_{\rm eff}$ to leading order in perturbation theory is
\begin{align*}  H_{\rm eff} =  \left( \begin{array}{cc} \boldsymbol{v}_{\boldsymbol{k}_{\rm W}}^{(n)\dagger}\boldsymbol{\mathcal{H}}\boldsymbol{v}_{\boldsymbol{k}_{\rm W}}^{(n)} & \boldsymbol{v}_{\boldsymbol{k}_{\rm W}}^{(n)\dagger}\boldsymbol{\mathcal{H}}\boldsymbol{v}_{\boldsymbol{k}_{\rm W}}^{(n+1)} \\ \boldsymbol{v}_{\boldsymbol{k}_{\rm W}}^{(n+1)\dagger}\boldsymbol{\mathcal{H}}\boldsymbol{v}_{\boldsymbol{k}_{\rm W}}^{(n)} & \boldsymbol{v}_{\boldsymbol{k}_{\rm W}}^{(n+1)\dagger}\boldsymbol{\mathcal{H}}\boldsymbol{v}_{\boldsymbol{k}_{\rm W}}^{(n+1)}   \end{array} \right) + \ldots.
\end{align*}
Corrections to higher order in perturbation theory involve mixing of states outside the two band subspace which are suppressed by the gap to the other bands.

For a Chern number $C=1$ Weyl point, the effective Hamiltonian takes the form  $H_{\rm eff} = \left( \omega_0 + \boldsymbol{A}^0 \cdot \delta \boldsymbol{k} \right) \sigma^0  + \delta k_i A^{ij} \sigma^j$. This Hamiltonian may be diagonalized unitarily to obtain the spectrum close to the touching point as well as the magnon eigenstates. 

We may reformulate the calculation of the neutron scattering cross section in terms of the two band subspace as follows. The neutron scattering intensity of mode $m$ at wavevector $\boldsymbol{k}$ may be written in the form $\langle \boldsymbol{k}, m \vert \hat{\mathsf{N}}_{\boldsymbol{k}} \vert \boldsymbol{k}, m\rangle$ where the ``scattering" operator $\hat{\mathsf{N}}_{ \boldsymbol{k}}$ in the sublattice basis is
\[ \boldsymbol{\mathsf{N}}_{ \boldsymbol{k}} = \left( \begin{array}{cc} \boldsymbol{\mathsf{N}}^{\dagger}_{1,ab} & \boldsymbol{\mathsf{N}}_{2,ab} \\ \boldsymbol{\mathsf{N}}^\dagger_{2,ab} & \boldsymbol{\mathsf{N}}_{1,ab}   \end{array} \right)
\]
with 
\begin{align}
\boldsymbol{\mathsf{N}}_{1,ab} & = P_{\alpha\beta}\left( \Gamma_a^{\alpha x}\Gamma_b^{\beta x} +  \Gamma_a^{\alpha y}\Gamma_b^{\beta y} - i \Gamma_a^{\alpha x}\Gamma_b^{\beta y} + i  \Gamma_a^{\alpha y}\Gamma_b^{\beta x} \right) \nonumber \\
\boldsymbol{\mathsf{N}}_{2,ab} & = P_{\alpha\beta}\left( \Gamma_a^{\alpha x}\Gamma_b^{\beta x} -  \Gamma_a^{\alpha y}\Gamma_b^{\beta y} + i \Gamma_a^{\alpha x}\Gamma_b^{\beta y} + i  \Gamma_a^{\alpha y}\Gamma_b^{\beta x} \right) \nonumber
\label{eqn:scattering}
\end{align}
and $P_{\alpha\beta}=\delta_{\alpha\beta}- \hat{k}_\alpha \hat{k}_\beta$ and where the states $\vert \boldsymbol{k}, m\rangle$ are represented by the $2N$ component column vector $\left( \left[ v_{\boldsymbol{k}} \right]_{n m+N}  \right)^{T}$.

We are interested in the vicinity of touching points detectable using neutron scattering. The transverse projector $P_{\alpha\beta}$ at such points is a much more slowly varying function of $\boldsymbol{k}$ than the magnon wavefunction, so we take it to be constant. We then project the scattering operator onto the eigenstates at the touching point the result being denoted $(\mathsf{N}_{\rm eff})_{m_1 m_2} $ for $m_1$ and $m_2=1,2$ which is defined up to a unitary transformation on the indices. This {\it effective scattering operator} takes the form $\boldsymbol{\mathsf{N}}_{\rm eff}=n_0 \mathbb{I} + n^i\sigma_i$ so the location of the touching point in momentum space picks out a direction $\hat{\boldsymbol{n}}$. For the above simple case of the 2D Heisenberg ferromagnet the intensity anisotropy vector $\hat{\boldsymbol{n}}$ is along the $\Gamma K$ direction passing through each Dirac point. Since the neutron scattering intensity is modulated from one Brillouin zone to another, the orientation of $\hat{\boldsymbol{n}}$ will generally be zone dependent.

The result of the above discussion is that the neutron scattering cross section in the vicinity of a touching point can be computed within the two band subspace. There are two ingredients: the effective neutron scattering operator that picks some particular direction, $\hat{\boldsymbol{n}}$, in momentum space. The second ingredient is the effective two band Hamiltonian that may be written $H_{\rm eff}=\boldsymbol{d}(\boldsymbol{k}-\boldsymbol{k}_{\rm W}) \cdot \boldsymbol{\sigma}$ in the vicinity of the touching point up to a momentum dependent shift whose eigenstates are momentum dependent spinors or pseudospins. The cross section is proportional to the pseudospin polarization along $\hat{\boldsymbol{n}}$. Weyl points have the property that the pseudospin texture winds around the touching point so the associated neutron scattering cross section winds also.

\begin{figure}[ht!]
\begin{center}
\includegraphics[width=0.6\columnwidth]{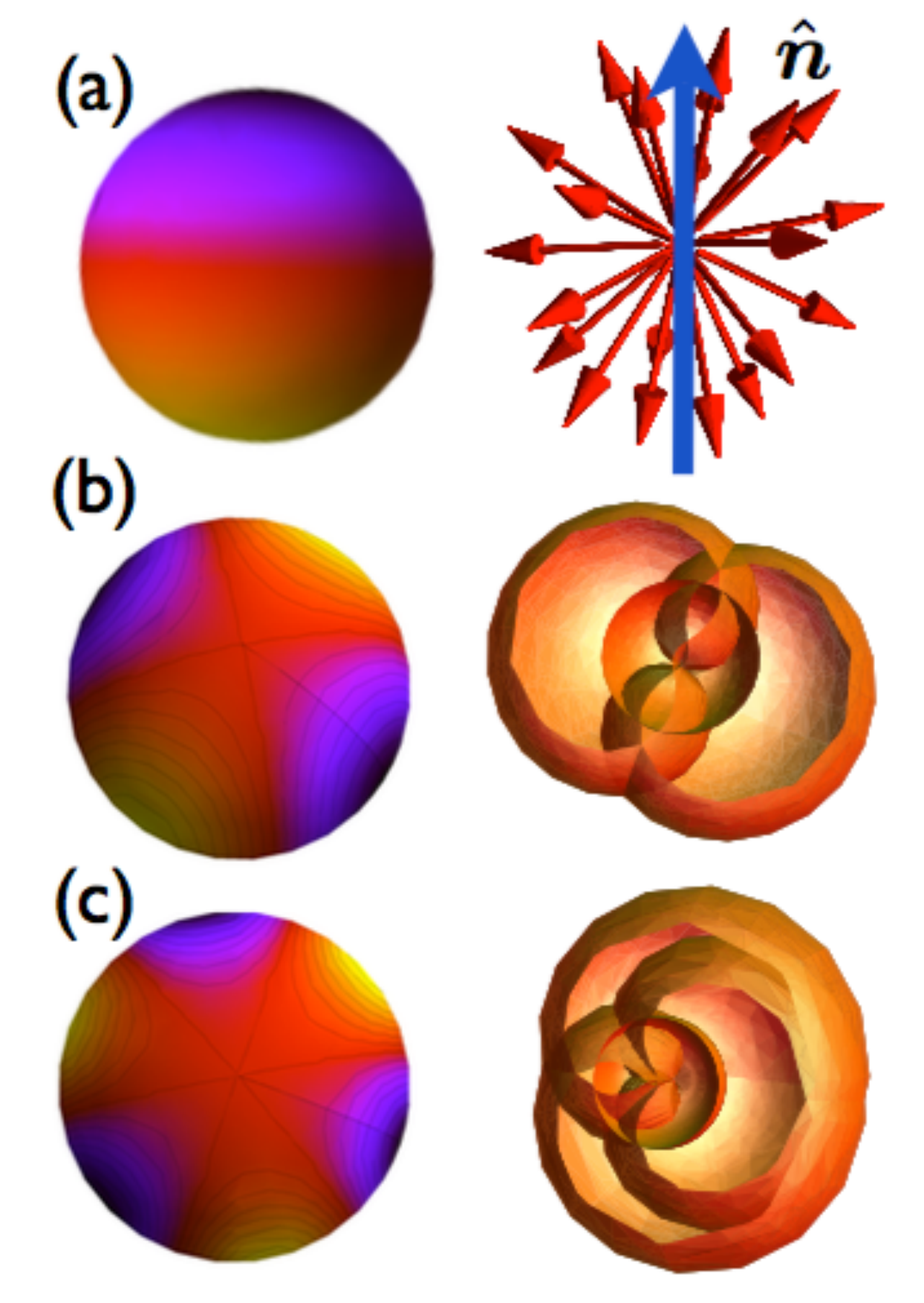}
\caption{Left-hand panels show the dynamical structure factor on a spherical surface in momentum space around Chern number $1$ (a), $2$ (b) and $3$ (c) Weyl points. The $C=1$ case has dipolar form and the anisotropy axis is vertical. In pseudospin space, the $\boldsymbol{d}$ vector has a hedgehog texture, sweeping out a spherical surface as $\boldsymbol{k}$ runs over the sphere, and the intensity shown corresponds to vertical $\hat{\boldsymbol{n}}$ vector. The right-hand plots (b) and (c) show surface plots of the $\boldsymbol{d}$ vector as a function of $\boldsymbol{k}$. The surface plots have been cut away to show the convolutions inside. The Chern number is equivalent to the number of times the surface wraps the origin. 
 }\label{fig:weylC123}
\end{center}
%\vspace{-25pt}
\end{figure}

%%%%%%%%%%%%%%%%%%%%%%%%%%%%%%%%%%%%%%%%%%%%%%%%%%%%%%%%%%%%%
%%%%%%%%%%%%%%%%%%%%%%%%%%%%%%%%%%%%%%%%%%%%%%%%%%%%%%%%%%%%%
%% GENERAL EXAMPLES
%%%%%%%%%%%%%%%%%%%%%%%%%%%%%%%%%%%%%%%%%%%%%%%%%%%%%%%%%%%%%
%%%%%%%%%%%%%%%%%%%%%%%%%%%%%%%%%%%%%%%%%%%%%%%%%%%%%%%%%%%%%

{\it General Examples} $-$  We first consider the isotropic Weyl point for which $\boldsymbol{d}(\delta\boldsymbol{k}\equiv\boldsymbol{k}-\boldsymbol{k}_{\rm W})=(\delta k_x,\delta k_y,\delta k_z)$. Then since the neutron scattering cross section amounts to a measurement of the pseudospin texture in momentum space along some direction, we find $1\pm\cos\theta$ with $\theta$ measured with respect to the $\hat{\boldsymbol{n}}$ direction. The resulting intensity over a spherical surface in momentum space around the origin is plotted in Fig.~\ref{fig:weylC123}(a). This is nothing other than the $Y_1^0$ spherical harmonic up to a constant. For the generalized Weyl point, the central singular point remains and the pattern of intensity is deformed around the anisotropy axis $\hat{\boldsymbol{n}}$. 

Our next examples are double and triple Weyl points - two band touching points associated with Chern number $\pm 2$ and $\pm 3$ respectively \cite{fang2012multi,yang2014classification}. An example of a double Weyl point has $\boldsymbol{d}(\delta \boldsymbol{k}) =\left(\delta k_x^2 - \delta  k_y^2, 2\delta k_x\delta  k_y, \delta k_z  \right)$ which has linear-quadratic dispersion. Fig.~\ref{fig:weylC123}(b) (right) shows the surface swept out by the $\boldsymbol{d}(\delta \boldsymbol{k})$ over a sphere taken around the Weyl point in momentum space. The surface completely wraps the origin twice hence the Chern number $C=2$. This convoluted surface is visible in the neutron scattering intensity provided $\hat{\boldsymbol{n}}$ does not lie along $z$. For example, if $\hat{\boldsymbol{n}}\parallel \hat{\boldsymbol{x}}$, the pattern is as shown in Fig.~\ref{fig:weylC123}(b) (left) so the intensity pattern about $C=2$ points can be predominantly $Y_2^m$. For $\hat{\boldsymbol{n}}\parallel \hat{\boldsymbol{z}}$ the Weyl point is singular with a single anisotropy axis as in Fig.~\ref{fig:weylC123}(a). A $C=3$ Chern number Weyl point is given by $\boldsymbol{d}(\delta\boldsymbol{k}) =\left( \delta k_x^3 - 3\delta k_x \delta k_y^2, \delta k_y^3 - 3\delta k_y \delta k_x^2, \delta k_z  \right)$ with linear-cubic dispersions. For $\hat{\boldsymbol{n}}\parallel \hat{\boldsymbol{x}}$ the neutron scattering intensity is as shown in Fig.~\ref{fig:weylC123}(c) (left). For completeness, if we consider a topologically trivial $C=0$ linear-quadratic touching point then $\boldsymbol{d}(\delta\boldsymbol{k}) =\left( \delta k_x^2 , \delta k_y^2, \delta k_z  \right)$. In this case, with $\hat{\boldsymbol{n}}\parallel \hat{\boldsymbol{x}}$ the intensity has minima along that axis and an equatorial band of constant intensity. 

%%%%%%%%%%%%%%%%%%%%%%%%%%%%%%%%%%%%%%%%%%%%%%%%%%%%%%%%%%%%%
%%%%%%%%%%%%%%%%%%%%%%%%%%%%%%%%%%%%%%%%%%%%%%%%%%%%%%%%%%%%%
%% LATTICE MODELS
%%%%%%%%%%%%%%%%%%%%%%%%%%%%%%%%%%%%%%%%%%%%%%%%%%%%%%%%%%%%%
%%%%%%%%%%%%%%%%%%%%%%%%%%%%%%%%%%%%%%%%%%%%%%%%%%%%%%%%%%%%%

Now we illustrate these ideas in the context of Weyl points appearing in frustrated magnets. We take several examples with different lattices, couplings and ground states to demonstrate the general nature of the scattering signatures around magnon Weyl points.

{\it Breathing Pyrochlore Antiferromagnet} $-$ The breathing pyrochlore lattice is an fcc lattice of identical ``up" tetrahedra joined by ``down" tetrahedra. We consider a model with moments placed on the breathing pyrochlore sites that are coupled through isotropic exchange with different magnitudes for the two types of tetrahedra and with a weak single ion easy plane anisotropy perpendicular to the local $\langle 111\rangle$ directions \cite{li2016weyl}:
\[
H_{\rm BP} = J \sum_{\langle i,j\rangle \in {\rm up}} \boldsymbol{\mathsf{S}}_i\cdot\boldsymbol{\mathsf{S}}_j + J' \sum_{\langle i,j\rangle \in {\rm down}} \boldsymbol{\mathsf{S}}_i\cdot\boldsymbol{\mathsf{S}}_j + \Delta \sum_{i} \left( \hat{\boldsymbol{z}}_i\cdot\boldsymbol{\mathsf{S}}_i \right)^2
\]
Quantum order-by-disorder selects a discrete six-fold set of states  \cite{li2016weyl}. Weyl points appear in the magnon spectrum about these ground states in the middle two bands (Fig.~\ref{fig:weyl}(a) (bottom)) at $(\pm k_{\rm W},0,0)$ and $(0,\pm k_{\rm W},0)$. Let us consider one of the former for which calculation of the full dynamical structure factor reveals the intensity to wind around the Weyl node with anisotropy axis along $k_x$ (Fig.~\ref{fig:weyl}(a) (top)). Projection onto the two band problem for $J=1$, $J'=0.6$ and $\Delta=0.2$ and $k_{\rm W}\approx (2\pi/a)*0.535$ yields an effective Hamiltonian with $\boldsymbol{d}(\delta\boldsymbol{k})=(A^{yx}\delta k_y+A^{zx}\delta k_z,A^{yy}\delta ky+A^{zy}\delta k_z,A^{xz}\delta k_x)$ so along $(\delta k_x,0,0)$ the $\boldsymbol{d}$ vector points in the $z$ direction. The effective neutron scattering operator direction is $\hat{\boldsymbol{n}}=\hat{\boldsymbol{z}}$ so the maximum and minimum of intensity are aligned along $(k_x,0,0)$ about the Weyl node in agreement with the full calculation.

{\it Pyrochlore Ferromagnet} $-$ We consider the weakly spin-orbit coupled ferromagnet with Dzyaloshinskii-Moriya exchange that is the appropriate leading order model to describe the magnon spectrum in the pyrochlore magnet Lu$_2$V$_2$O$_7$ \cite{onose2010observation,mena2014spin}. The Hamiltonian is
\[ H_{\rm PF}=-\vert J \vert\sum_{\langle i,j\rangle} \boldsymbol{\mathsf{S}}_i\cdot\boldsymbol{\mathsf{S}}_j + \sum_{\langle i,j\rangle} \boldsymbol{D}_{ij}\cdot\left( \boldsymbol{\mathsf{S}}_i\times\boldsymbol{\mathsf{S}}_j\right). \]
where the $\boldsymbol{D}_{ij}$ directions are fixed by the lattice symmetries \cite{SM}. The ground states are the rotationally symmetric collinear ferromagnetic states. When $D\equiv \vert D_{ij}\vert\neq 0$, the magnon spectrum contains a pair of Weyl points between the middle two bands \cite{mook2016tunable,su2017magnonic,SM}. The Weyl points lie along the magnetization direction when this is along high symmetry lines and the anisotropy axis in the Weyl point dynamical structure factor lies along the same direction - the case of moments polarized along $[111]$ is shown in Fig.~\ref{fig:weyl}(b) which illustrates the intensity winding around the Weyl node. For this case, in the language of the effective Hamiltonian, the $\boldsymbol{d}(\delta \boldsymbol{k})$ vector lies along $\hat{\boldsymbol{n}}$ when $\boldsymbol{k}=\boldsymbol{k}_{\rm W}+(\delta k,\delta k,\delta k)$ as may be calculated directly by projecting onto the Weyl point \cite{SM}.

{\it Stacked Honeycomb Ferromagnet} $-$ The honeycomb ferromagnet, discussed above, when stacked has Dirac line nodes in the $(00L)$ direction. When the second in-plane neighbors are coupled by Dzyaloshinskii-Moriya exchange, the Dirac point in the 2D ferromagnet is gapped out leading to Chern bands with Chern number $\pm 1$. When the layers are coupled the modes disperse in the $(00L)$ direction and particular couplings lead to band inversion at Weyl points \cite{su2017chiral,SM}. Again, projection onto the two band problem in the vicinity of the Weyl node at $(k_x,0,0)$ reveals the anisotropy direction is in the $(1,0,0)$ direction (Fig.~\ref{fig:weyl}(c)) \cite{SM}.

In conclusion, we have calculated the neutron scattering intensity from sharp magnon touching points and lines showing that it is a measure of the polarization in some fixed direction of a spinor texture in momentum space. For magnon Dirac points in 2D and Weyl points in 3D the nodes are singular points with a general  and characteristic intensity pattern in their vicinity. We expect the findings presented here to be applicable to topological magnon band structures in spin-orbit coupled magnets. More concretely, the presence of Weyl points leads to ``intensity arcs" in constant energy maps as illustrated in Fig.~\ref{fig:dirac_weyl}(b). The direction of $\hat{\boldsymbol{n}}$ may vary between Weyl points in different Brillouin zones providing different snapshots of the pseudospin texture in momentum space. In addition, one may use a magnetic field to tune the position of Weyl points and sweeping out a range of $\hat{\boldsymbol{n}}$ allowing one to map out the $\boldsymbol{d}(\delta \boldsymbol{k})$ vector at the Weyl point. 

\begin{acknowledgments}
P.A.M. would like to thank Bitan Roy for drawing his attention to work on $C=2,3$ points and Robert-Jan Slager and Onur Erten for discussions. S.S. acknowledges the summer undergraduate internship programme at the MPI PKS during which a part of this work was conducted.  
\end{acknowledgments}

\bibliography{references}
\end{document}